\documentclass[prb,twocolumn,groupeaddress]{revtex4-1}

\usepackage{amsmath}
\usepackage{amssymb}
\usepackage{xspace}
\usepackage{graphicx}
\usepackage{grffile}
\usepackage{nicefrac}
\usepackage{color}

\graphicspath{{figs/}}

\def\bk{{\bf k}}
\def\bR{{\bf R}}
\begin{document}

\title{Hidden Mott insulator in metallic PdCrO$_2$}

\author{Frank Lechermann}
\affiliation{I. Institut f{\"u}r Theoretische Physik, Universit{\"a}t Hamburg, 
D-20355 Hamburg, Germany}

\begin{abstract}
There has been a long-standing debate on the coexistence between itinerant electrons 
and localized spins in the PdCrO$_2$ delafossite. By means of the charge
self-consistent combination of density functional theory and dynamical mean-field 
theory, it is corroborated that despite overall remarkable metallic response, the 
CrO$_2$ layers are indeed Mott insulating as suggested by previous experimental
work. The resulting ${\bf k}$-resolved spectral function in the paramagnetic phase is 
in excellent agreement with available photoemission data. Subtle coupling between the 
itinerant and Mott-localized degrees of freedom is revealed. Different doping 
scenarios are simulated in order to manipulate the electronic states within the 
inert layers. In particular, oxygen vacancies prove effective in turning the hidden 
Mott insulator into a strongly correlated itinerant subsystem. The present results 
may open a new venue in research on quantum materials beyond canonical 
classification schemes.
\end{abstract}


\maketitle

\section{Introduction}
The discrimination of distinct electronic phases is a standard paradigm
in condensed matter physics. Fermi-liquid metals, band insulators and Mott insulators 
form the canonical ones. Several doped Mott insulators, for instance, belong to a 
more complex class, but still, material characterization usually focuses on the global 
aspects given by transport and magnetism.
Recently, oxide heterostructures challenged this viewpoint, by establishing
rather different electronic states in selected real-space regions in a controlled
way. There, e.g. conducting interface regions are replaced by highly Mott-insulating 
areas further away from the interface~\cite{jac14,lec15,lec17}. 
The question arises if such coexistence of itinerancy and Mott localization can 
also be found in natural materials.

\begin{figure}[t]
\begin{center}
\includegraphics*[width=7cm]{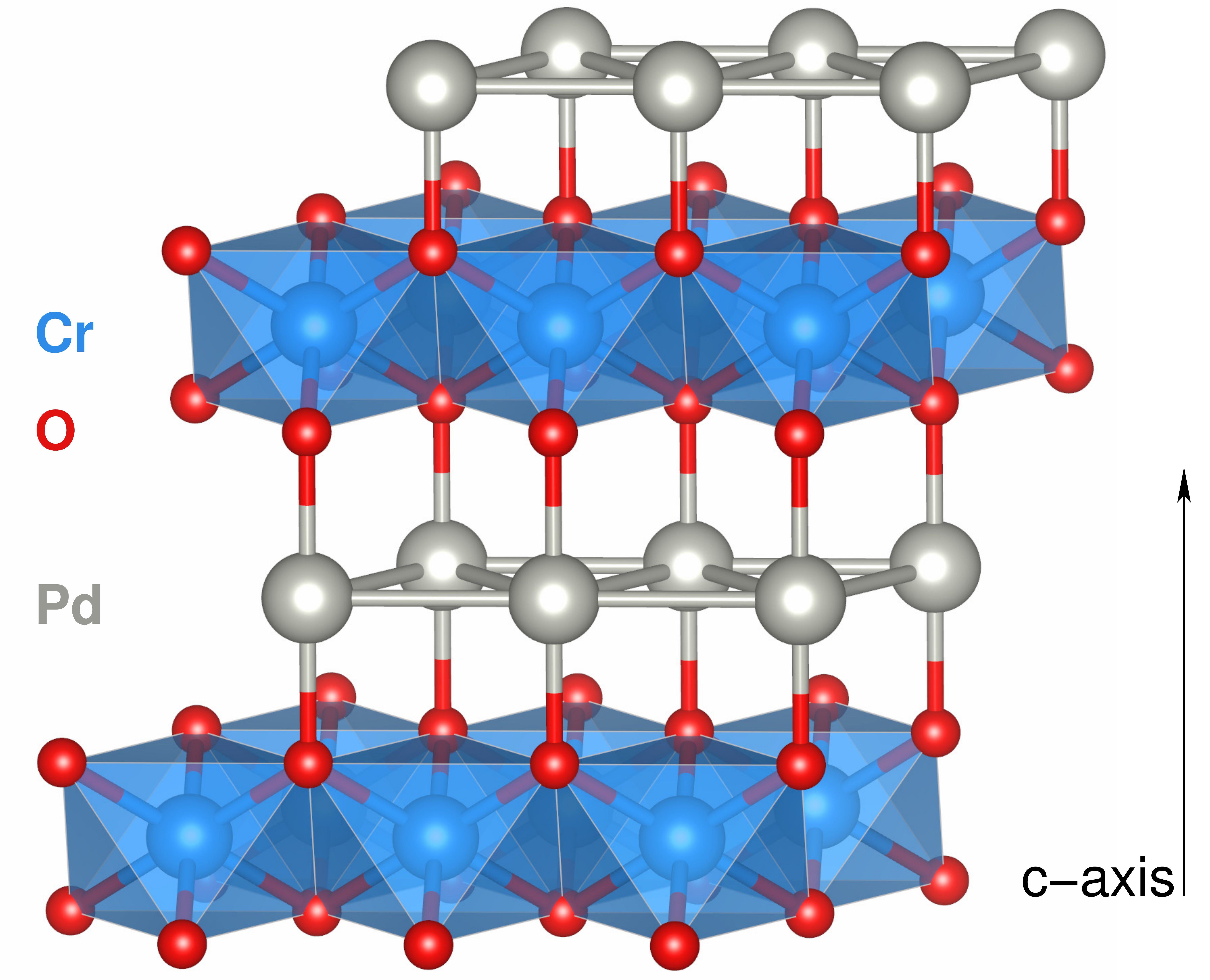}
\end{center}
\caption{(color online)
Delafossite $AB$O$_2$ crystal structure exemplified for PdCrO$_2$ with 
Pd (grey), Cr (blue) and O (red).
\label{fig:struc}}
\end{figure}

In that respect, the delafossite compound PdCrO$_2$~\cite{tak09,tak10} provides an 
intriguing test case. 
Selected $AB$O$_2$ delafossites~\cite{sha71-1,*sha71-2,*sha71-3}, where $A$ and $B$ 
are different transition-metal (TM) atoms, such as PdCoO$_2$, PtCoO$_2$ and PdCrO$_2$ 
are surprisingly high conductive (see e.g.~Refs.~\onlinecite{mac17,dao17} for recent
reviews). The room-temperature in-plane resistivity of 
PdCoO$_2$ amounts to 2.6 $\mu\Omega$cm~\cite{hic12}, rendering it the most-conductive 
oxide in this temperature $(T)$ range. The crystal structure displayed in 
Fig.~\ref{fig:struc} shows the delafossite stacking of triangular $A$-atom planes and 
trigonal $B$O$_2$ planes along the $c$-axis exemplified for PdCrO$_2$. As a key fact, 
the TM environments strongly differ for $A$ and $B$ sites. The $B$ sites
are in a common octahedral coordination with layered octahedra, giving rise to 
trigonal $d$-shell states of $t_{2g}=\{a_{1g},e_g'\}$ character at low energy and of 
$e_g$ nature at high energy. 
On the other hand, the $A$ sites are linked to oxygens in a dumbbell-like
position up and below, eventually leading to a more unusal $d$-state 
hierachy~\cite{eye08}. Experimental data strongly favors the picture of rather inert 
$B$O$_2$ layers in terms of itinerancy. From a general formal-oxidation analysis, the 
valences $A^{+1}B^{3+}$O$^{2-}_2$ describe cobalt in (Pd,Pt)CoO$_2$ compounds as 
Co$^{3+}(3d^6)$ with a low-spin closed-$t_{2g}$ subshell structure. This resembles the 
Co state in related NaCoO$_2$, the band-insulating layered rock salt~\cite{taka03} 
with both Na and Co in a trigonal position. Though already the interplay between the 
itinerant $4d$ electrons and the inert $3d^6$ configuration of cobalt raises 
interesting questions, the case of PdCrO$_2$ appears even more intriguing. 
Formal low-spin Cr$^{3+}(3d^3)$ has a half-filled $t_{2g}$ shell, with expected sizable 
local Coulomb interaction. Mott criticality in the CrO$_2$ layers becomes possible and 
several experiments indeed point towards localized Cr$^{3+}$ 
spins~\cite{tak09,noh14,hic15} in contact with itinerant Pd$(4d)$ electrons. 
Especially the x-ray absorption data of Noh {\sl et al.} provides evidence for a 
Cr$^{3+}$(-like) oxidation state. Yet a fully conclusive proof for complete absence 
of itinerancy in the CrO$_2$ layers is still missing. Note also finally, magnetic 
ordering in a 120$^\circ$ spin structure is revealed~\cite{mek95} below 
$T_{\rm N}=37.5\,$K.
%%%%%%%%%%%%%%%%%%%%%%%%%%%%%%%%%%%%%%%%%%%%%%%%%%%%%%%%%%%%%%%%%%%%%%%%%%%%%%%%%%%%%%%%%
%%%%%%%%%%%%%%%%%%%%%%%%%%%%%%%%%%%%%%%%%%%%%%%%%%%%%%%%%%%%%%%%%%%%%%%%%%%%%%%%%%%%%%%%%

In this work, a detailed study of the correlated electronic structure of paramagnetic
PdCrO$_2$ is presented. As suggested from experiment, it indeed confirms the 
Mott-insulating character of the CrO$_2$ layers from first-principles many-body theory. 
Due to its concealed nature in the overall very good metallic system, one may classify 
this real-space-selective insulating regime as a 'hidden Mott insulator'. The 
${\bf k}$-resolved interacting spectral function differs substantially from the effective 
single-particle band structure obtained within conventional density functional theory 
(DFT). As the spectral weight of 
low-energy Cr-like DFT bands is shifted to higher energies with correlations, the 
Fermi level gets positioned in a one-band Pd-dominated quasiparticle (QP) dispersion. 
Significant coupling between the localized Cr electrons and the itinerant Pd
electrons is discovered from analyzing the electronic self-energy. Different
doping scenarios are employed to perturb the hidden state, eventually
rendering the CrO$_2$ layers itinerant.

\section{Theoretical Approach}
The charge self-consistent combination of density functional theory and 
dynamical mean-field theory (DMFT) is employed~\cite{sav01,pou07,gri12}.
For the DFT part, I use the mixed-basis pseudopotential method~\cite{lou79,mbpp_code}, 
based on norm-conserving pseudopotentials with a combined basis of localized functions 
and plane waves. For the exchange-correlation part in DFT, the generalized-gradient 
approximation in form of the PBE functional~\cite{per96} is utilized. 
Within the mixed basis, localized functions for Cr($3d$) and Pd($4d$) states as well 
as for O($2s$) and O($2p$) are used in order to reduce the plane-wave energy cutoff.
The correlated subspace consists of the effective Cr($3d$) Wannier-like 
functions as obtained from the projected-local-orbital formalism~\cite{ama08,ani05}, 
using as projection functions the linear combinations of atomic $3d$ orbitals, 
diagonalizing the Cr($3d$) orbital-density matrix. I use a five-orbital
Slater-Kanamori Hubbard Hamiltonian in the correlated subspace, parametrized by
a Hubbard $U=3$\,eV and a Hund's exchange $J_{\rm H}=0.7$\,eV. The latter values
represent a proper choice for chromium oxides~\cite{kor98} and for 
akin layered CoO$_2$ compounds~\cite{lec09}. Since the relevant $d$-states of 
palladium are of $4d$ character and are furthermore close to complete filling in 
PdCrO$_2$, the effect of explicit local Coulomb interactions on the Pd site may
be safely neglected for examining the qualitative key physics.
The single-site DMFT impurity problems in stoichiometric and doped PdCrO$_2$ are 
solved by the continuous-time quantum Monte Carlo 
scheme~\cite{rub05,wer06} as implemented in the TRIQS package~\cite{par15,set16}. 
A double-counting correction of fully-localized type~\cite{ani93} is applied. 
To obtain the spectral information, analytical continuation from Matsubara space via 
the maximum-entropy method as well as the Pad{\'e} method is performed. All DFT+DMFT
calculations are conducted by setting the system temperature to $T=290$\,K.
Paramagnetism is assumed in all those computations.

Experimental lattice parameters~\cite{sha71-1,*sha71-2,*sha71-3} $a=2.930$\,\AA\, 
and $c=18.097$\,\AA\, are used. 
The internal degree of freedom $z$, governing the oxygen distance to 
the Pd plane is obtained from DFT structural optimization, reading $z=0.1101$ for the 
stoichiometric compound.

\section{Results}
\subsection{DFT electronic structure}
%%%%%%%%%%%%%%%%%%%%%%%%%%%%%%%%%%%%%%%%%%%%%%%%%%%%%%%%%%%%%%%%%%%%%%%%%%%%%%%%%%%%%%%%%
\begin{figure}[t]
\begin{center}
(a)\hspace*{-0.4cm}\includegraphics*[width=8.25cm]{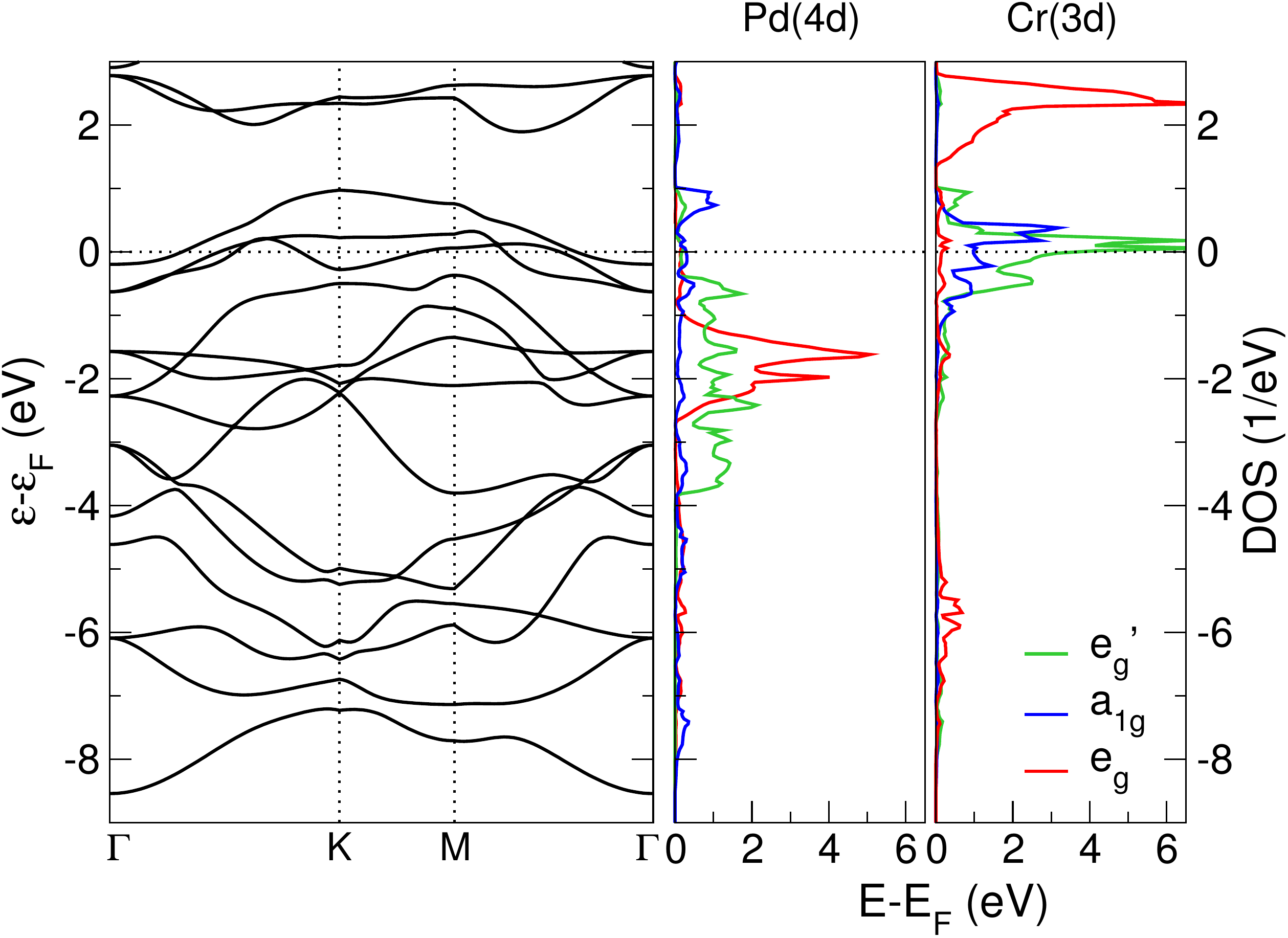}\\[0.2cm]
(b)\hspace*{-0.4cm}\includegraphics*[width=8.25cm]{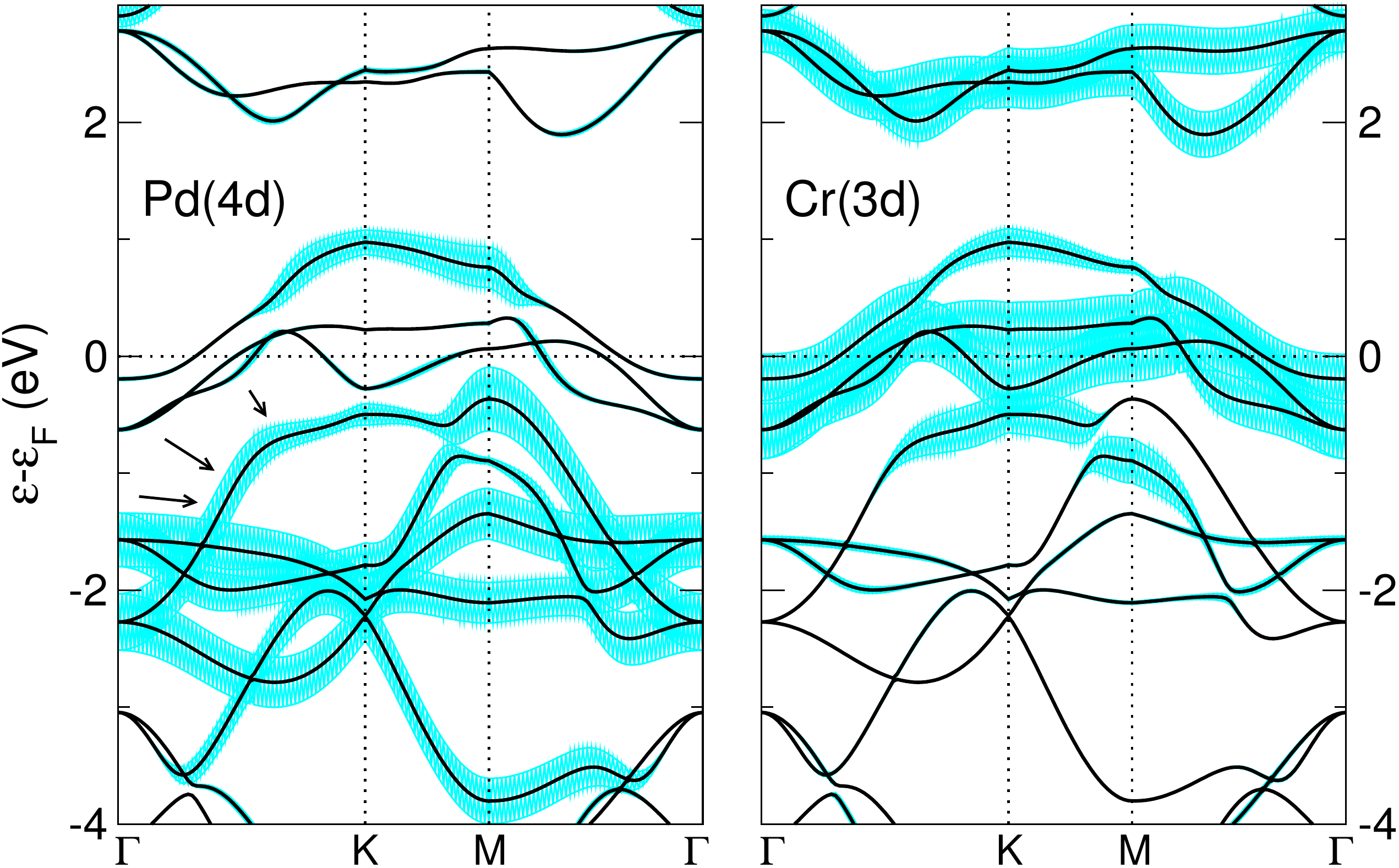}\\[0.4cm]
(c)\includegraphics*[width=8cm]{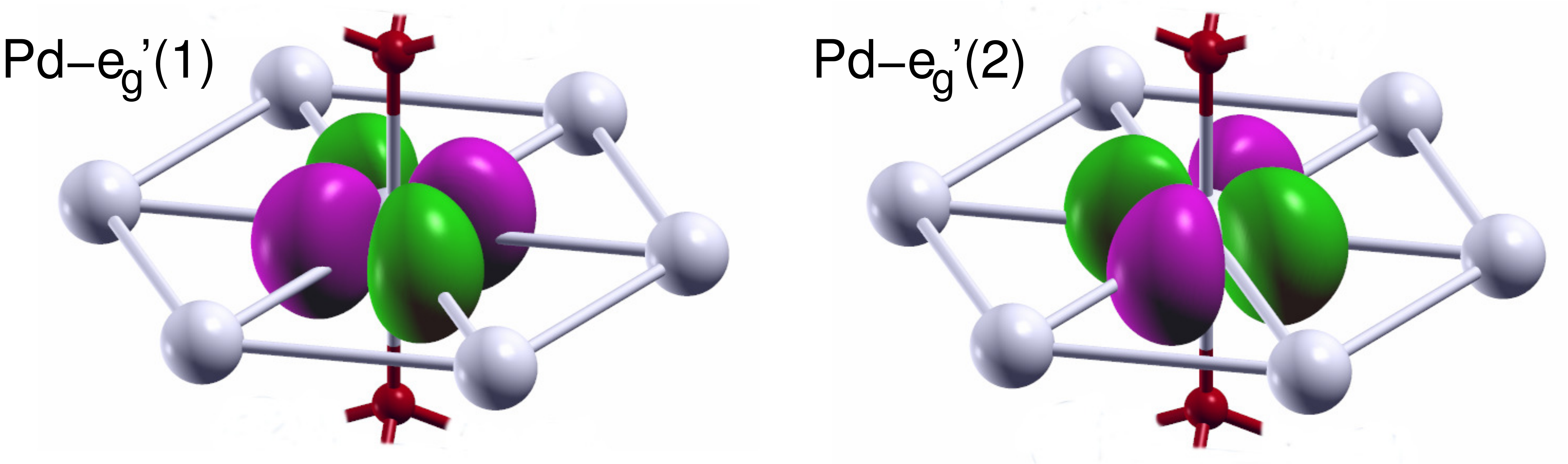}\\[0.2cm]
\end{center}
\caption{(color online)
Bandstructure and density of states (DOS) from DFT.
(a) Bands within the $k_z=0$ plane along high-symmetry directions (left) and 
local density of states for Pd and Cr (right). Note that the Pd-$a_{1g}$ 
orbital point towards the apical oxygen and the Pd-$e_{g}'$ marks strictly
in-plane orbitals. 
(b) Pd($4d$) (left) and Cr($3d$) (right) character on the TM-dominated bands.
Small arrows in left part point to the cPd band (see text).
(c) In-plane Pd-$e_g'$ orbitals, dominating the band just below the low-energy 
Cr-$t_{2g}$ manifold.}
\label{fig:dftbasic}
\end{figure}
%%%%%%%%%%%%%%%%%%%%%%%%%%%%%%%%%%%%%%%%%%%%%%%%%%%%%%%%%%%%%%%%%%%%%%%%%%%%%%%%%%%%%%%%%
Within Kohn-Sham density functional theory, the main results of 
characterizing nonmagnetic PdCrO$_2$ are displayed in Fig.~\ref{fig:dftbasic}. 
The relevant band structure and density of states (DOS) encompassing Pd$(4d)$, Cr$(3d)$ 
and O$(2p)$ states spans an energy window of $[-8.5,3]$\,eV 
(cf. Fig.~\ref{fig:dftbasic}a). 
The Cr levels are higher in energy than the Pd ones, rendering the
latter not too far from complete filling. At the Fermi level, a separated manifold of 
three bands marks a dispersion of width 1.6\,eV. These bands are nearly
exclusively of Cr$(3d)$ character as also visible from the 'fatband' plot of 
Fig.~\ref{fig:dftbasic}b, which visualizes the underlying Pd$(4d)$ and Cr$(3d)$ 
contribution to each band.

From a local-orbital analysis (see e.g. also Ref.~\onlinecite{eye08} for the orbital
setting), the trigonal 
splitting at the Cr site with octahedral coordination marks the given low-energy bands 
as being of dominant $t_{2g}$ kind. It  consists of $a_{1g}$ and doubly-degenerate $e_g'$ 
orbitals. Note that the Cr-$a_{1g}$ orbital points along the $c$-axis, while the 
Cr-$e_g'$ orbitals point in direction between the oxygen ligands and are inclined
to the vertical axis. The remaining Cr-$e_g$ orbitals of higher energy are directed 
towards the oxygens. Because of the different site symmetry, the Pd-$a_{1g}$ 
orbital points to the apical oxygens and therefore does not mark a low-energy orbital 
as in the Cr case.

Remarkably, and as pointed out already in Ref.~\onlinecite{mac17}, the just 
described DFT-based low-energy electronic structure disagrees
completely with angle-resolved photoemission spectroscopy (ARPES) 
experiments~\cite{sob13,noh14}. There, only a single QP band crosses the Fermi level,
which also from quantum-oscillation experiments~\cite{ok13,hic15} is designated as 
being dominantly of Pd character. The obvious solution to this discrepancy is given
by the fact that the Cr states are effectively localized due to electronic correlations
based on sizable local Coulomb interactions. In other words,
the CrO$_2$ layers become Mott insulating, while the Fermi level of the still overall
metallic system shifts into a Pd-dominated QP band. From Fig.~\ref{fig:dftbasic}b,
it appears obvious to which principle band it comes down to: the band just below the
Cr-based low-energy threefold of bands, which is of strong Pd$(4d)$ character, has
to take charge. Importantly when starting from $\Gamma$, that band carries dominantly 
Pd-$e_g'$ weight (cf.  Fig.~\ref{fig:dftbasic}a,b), representing twofold in-plane 
orbitals (see Fig.~\ref{fig:dftbasic}c). Closer to the $K$ point, Cr and Pd-$a_{1g}$ 
contribute strongest to that band. In the following, I will call this relevant band
the 'conducting Pd (cPd) band'. Note that Pd adds also some 
angular-momentum contribution from $s$ and $p$ over the relevant energy range, 
but the corresponding magnitude is always about 5-10 times smaller than from $d$ 
flavour.

Regarding the general spectral properties, some agreement with experiment can be 
achieved within spin-polarized DFT calculations for a magnetically ordered 
system~\cite{ong12,sob13,noh14,bil15}. However in the next section it will be shown 
that agreement with experiment is not truly associated with invoking 
magnetic order, but by correctly including the physics of generic electron 
correlations beyond conventional DFT.

\subsection{Correlated electronic structure of PdCrO$_2$}
\subsubsection{Main spectral properties}
%%%%%%%%%%%%%%%%%%%%%%%%%%%%%%%%%%%%%%%%%%%%%%%%%%%%%%%%%%%%%%%%%%%%%%%%%%%%%%%%%%%%%%%%%
\begin{figure}[b]
%\begin{center}
(a)\includegraphics*[width=8.25cm]{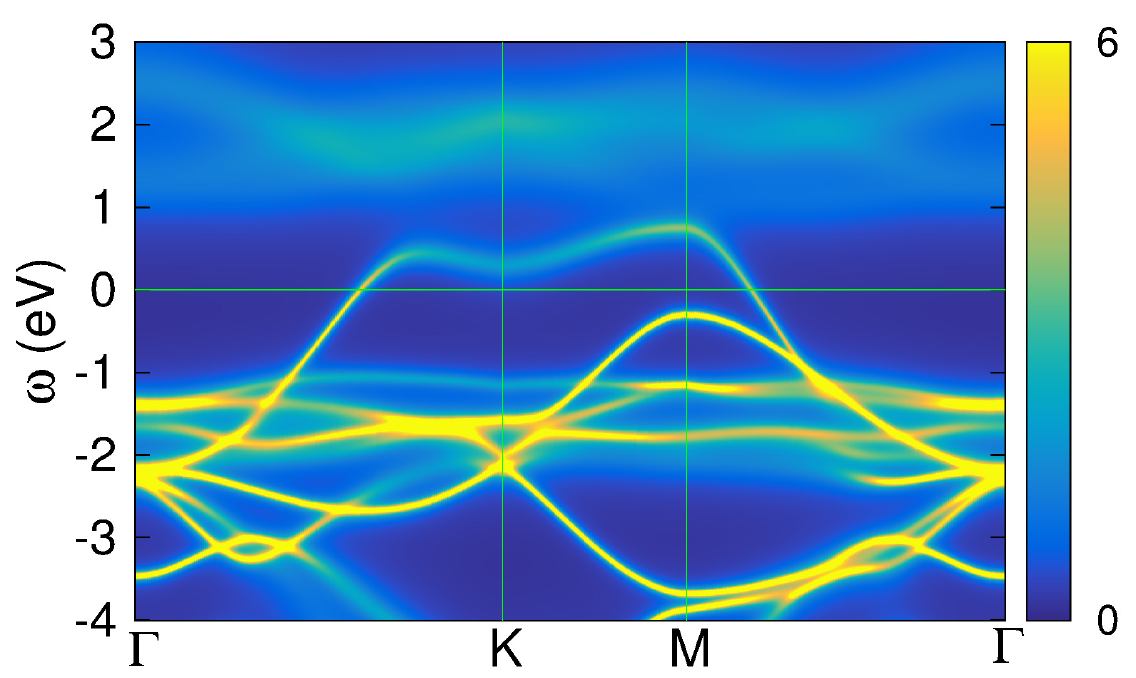}\\[0.1cm]
\includegraphics*[width=4.2cm]{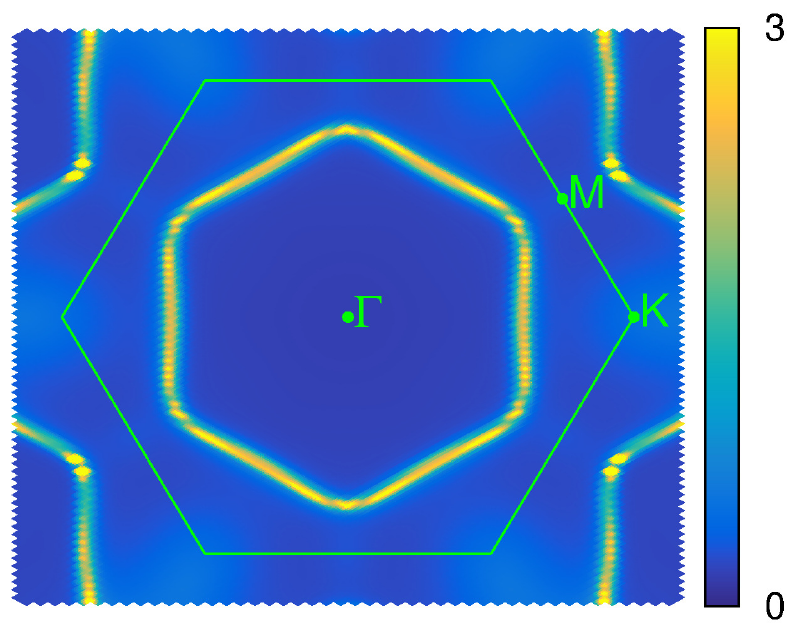}\hspace*{0.1cm}
\includegraphics*[width=4.2cm]{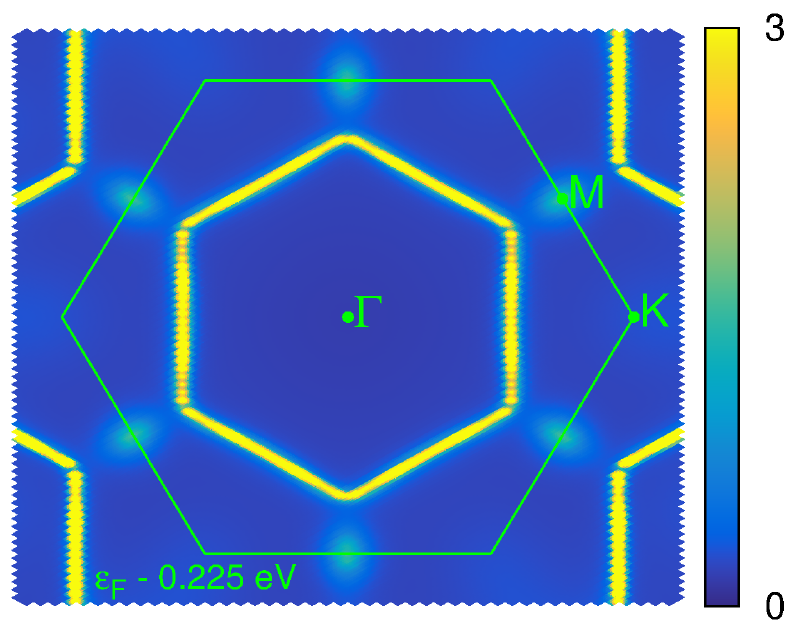}\\
\raggedright(b)\\[0.2cm]
(c)\hspace*{-0.4cm}\includegraphics*[width=8.25cm]{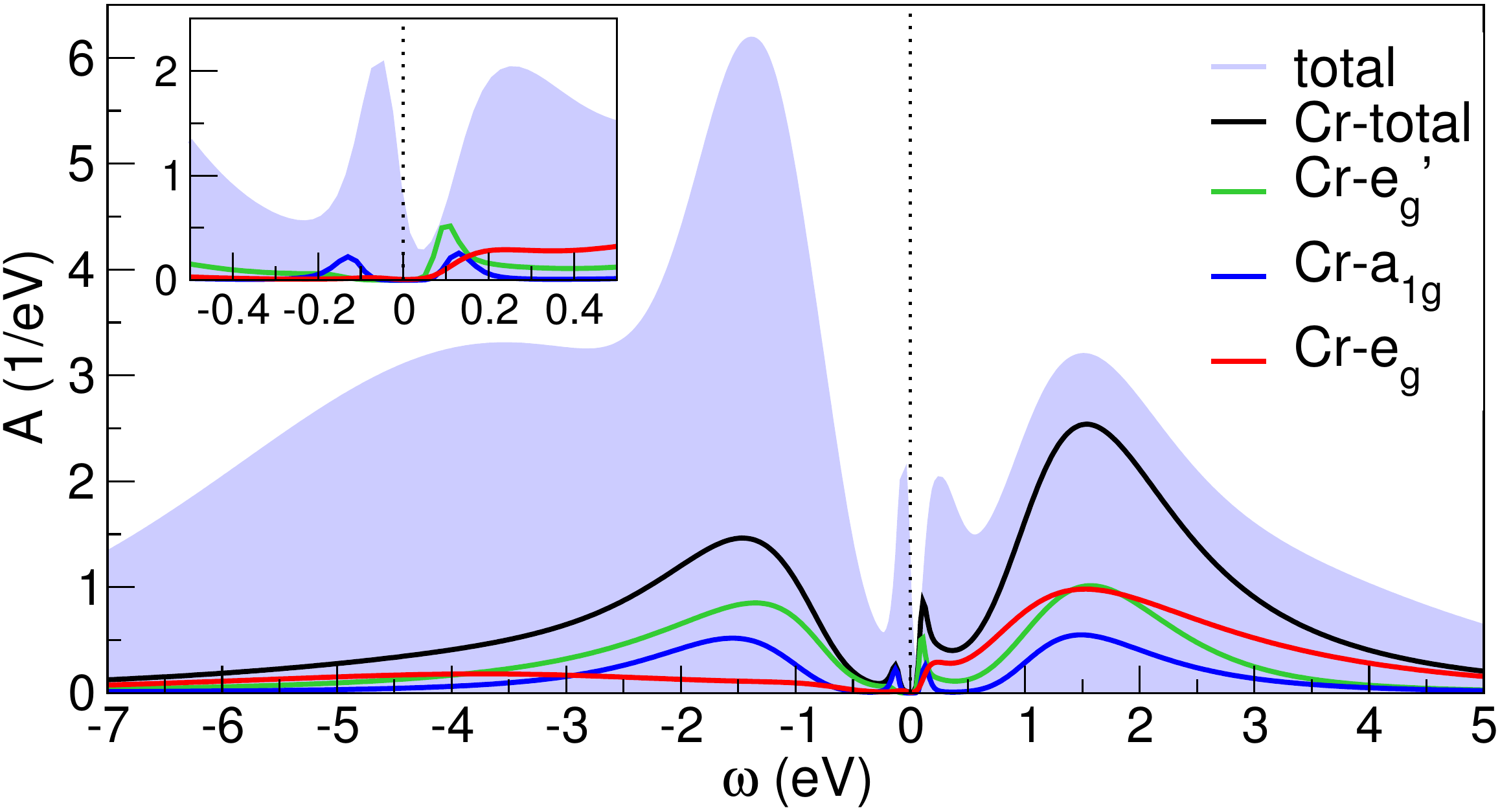}
%\end{center}
\caption{(color online) Spectral data from DFT+DMFT.
(a) Spectral function $A({\bf k},\omega)$ along high-symmetry lines in the $k_z=0$
plane of reciprocal space.
(b) Constant-energy surfaces within the first Brillouin zone (green hexagon): 
Fermi surface $\omega=\varepsilon^{\hfill}_{\rm F}=0$ (left) and 
$\omega=\varepsilon^{\hfill}_{\rm F}-0.225$\,eV.
(c) ${\bf k}$-integrated spectral function $A(\omega)$ with orbital-resolved 
local Cr part. Inset: blow-up around Fermi level.}
\label{fig:spectra}
\end{figure}
%%%%%%%%%%%%%%%%%%%%%%%%%%%%%%%%%%%%%%%%%%%%%%%%%%%%%%%%%%%%%%%%%%%%%%%%%%%%%%%%%%%%%%%%%
Figure~\ref{fig:spectra} summarizes the spectral information that is obtained with 
charge self-consistent DFT+DMFT. The ${\bf k}$-resolved spectral function 
$A({\bf k},\omega)$
plotted in Fig.~\ref{fig:spectra}a along high-symmetry lines in the $k_z=0$ plane of
reciprocal space exhibits the expected matching with experimental findings. A single
QP band crosses the Fermi energy $\varepsilon^{\hfill}_{\rm F}$ along $\Gamma$-$K$ and
$\Gamma$-$M$. Comparison of the band labelling with the original DFT bandstructure
easily verifies it as indeed stemming dominantly from the cPd band. The Fermi wavevector
with a fraction of 0.61 along the complete $\Gamma$-$K$ path, as well as the central-peak
energy of $-0.29$\,eV of the occupied hole pocket at the $M$-point are in very good
agreement with ARPES data~\cite{sob13,noh14}. While the occupied Pd bands away from the
Fermi level show mostly proper coherence, the original Cr-$e_g$ bands above 
$\varepsilon^{\hfill}_{\rm F}$ are now strongly incoherent. Importantly, the former
Cr-$t_{2g}$ bands at low-energy have fully disintegrated. The Fermi surface in
Fig.~\ref{fig:spectra}b agrees with the single-sheet hexagon centered around $\Gamma$ 
from experiment~\cite{sob13}, and also the constant-energy cut somewhat below 
$\varepsilon^{\hfill}_{\rm F}$ reveals again the same pocket structure around the 
symmetry-equivalent $M$-points.

Of course, the Cr spectral weight has not vanished into thin air. As shown from a plot
of the ${\bf k}$-integrated total spectral function and the local chromium part in 
Fig.~\ref{fig:spectra}c, the Cr spectral weight is broadly distributed over energies 
away from the Fermi level, with dispersive parts mingling with the Pd bands. 
Right at $\varepsilon^{\hfill}_{\rm F}$ there is zero spectral weight, i.e. Cr is 
charge gapped and
the CrO$_2$ layer indeed Mott insulating. Due to the overall metallicity of the 
system, this may be called a 'hidden Mott insulator' in a real-space selective region
of the material. From theory, such Mott states in itinerant systems have e.g. 
already be found in oxide heterostructures~\cite{lec15,lec17}.
The 'lower Hubbard band' of hidden-Mott Cr extends over the
Pd-dominated energy region and becomes therefore rather invisible as a distinct
excitation in $A({\bf k},\omega)$ (cf. Fig.~\ref{fig:spectra}a). The orbital-resolved
Cr occupations from DFT+DMFT read $n=\{n_{e_g'},n_{a_{1g}},n_{e_g}\}=\{2.07,1.02,1.06\}$
with a total electron count of $n_{\rm tot}=4.15$. At first glance, this seems to
disagree with the straightforward picture of Cr$^{3+}$ with spin $S=3/2$ that is
usually put forward from basic considerations. However note that the Wannier-based
Cr-$e_g$ spectral weight of one electron is very flat in the occupied region, 
marking the broad hybridization with its surrounding. Thus, the more localized 
$t_{2g}$ electrons shall count most when it comes to local properties and therefrom 
Cr$^{3+}$ with $S=S_{t_{2g}}\approx 3/2$ indeed matches expectations. In this
respect, note that Pd is nominally in the '+1' oxidation state, which is usually
interpreted as a $4d^9$ configuration. 

Finally, one may use the opportunity to roughly compare CrO$_2$ in PdCrO$_2$ with 
well-known CrO$_2$ in the rutile structure~\cite{kor98,bis17} in terms of the
correlation strength. In the former variation, correlations are indeed 
expected stronger from the Cr$^{3+}$ oxidation state, compared to
Cr$^{4+}$ in rutile CrO$_2$. It is known from basic model-Hamiltonian studies 
(e.g. Ref.~\onlinecite{pie18}) that the Mott transition with respect to the strength of 
the local Hubbard $U$ happens more quickly at half filling. Moreover, the 
quasi-twodimensional structure of CrO$_2$ in PdCrO$_2$ should further sustain 
tendencies towards correlation effects.

\subsubsection{Coupling of localized and itinerant electrons}
So far, the advanced electronic structure theory beyond DFT was mainly utilized to
achieve agreement with available experimental findings. However, it reveals also
important new insight into the crucial coupling of itinerant electrons with dominant
Pd character and localized Cr electrons~\cite{ok13,gla14}. 

Traditionally, such couplings are known from Kondo physics, namely in an incoherent
way associated with a Kondo impurity and in the coherent way on a Kondo lattice. The
latter may apply to heavy-fermion compunds, where a regular lattice of localized 
$f$-electron spins interacts with a surrounding Fermi sea~\cite{mot74,don77}. 
Key competition in such a system is between the antiferromagnetic 
Rudermann-Kittel-Kasuya-Yosida (RKKY) interaction trying to order the spins and the 
Kondo effect trying to screen the spins. Seemingly, the present PdCrO$_2$ problem 
exhibits some similarities to the basic architecture of the Kondo-lattice problem, 
but importantly, the present localized spins do not result from atomic physics of 
partially-filled inner $f$-shells. Instead, the Cr spins originate from a Mott 
mechanism that suppresses the hopping via strong local Coulomb repulsion. Little 
formal details are known about this regime of itinerant-localized electron coupling.
%%%%%%%%%%%%%%%%%%%%%%%%%%%%%%%%%%%%%%%%%%%%%%%%%%%%%%%%%%%%%%%%%%%%%%%%%%%%%%%%%%%%%%%%%
\begin{figure}[t]
\includegraphics*[width=8.5cm]{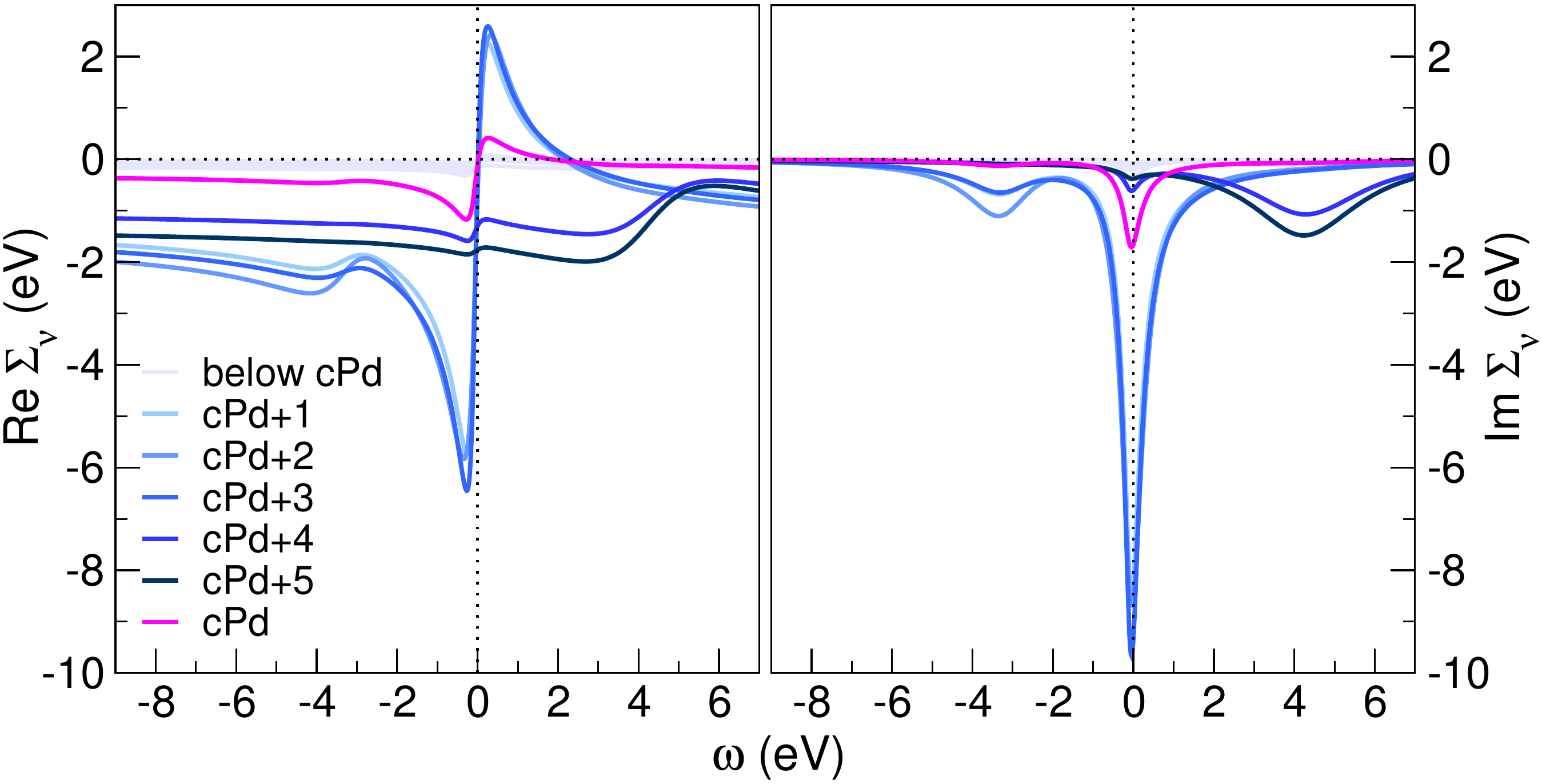}
\caption{(color online) Diagonal elements of the two-dimensional self-energy 
$\Sigma^{\rm 2D}_{\nu\nu'}(\omega)$ in the band basis. Left: real part, right: 
imaginary part. The self-energy on the cPd band is highlighted (magenta).}
\label{fig:sigma}
\end{figure}
%%%%%%%%%%%%%%%%%%%%%%%%%%%%%%%%%%%%%%%%%%%%%%%%%%%%%%%%%%%%%%%%%%%%%%%%%%%%%%%%%%%%%%%%%

As already shown in Fig.~\ref{fig:spectra}c, there is some minor Cr spectral weight
appearing at low energy close to $\varepsilon^{\hfill}_{\rm F}$, which is a hint
for a coupling to the itinerant Pd electrons. Furthermore on the DFT level, 
there is substantial Cr$(3d)$ weight on the cPd band around the $K$-point 
(see Fig.~\ref{fig:dftbasic}b). It is therefore expected that the itinerant Pd electrons 
carry some self-energy (or 'heaviness') from scattering with the highly correlated Cr 
electrons. This can be made quantiative via the self-energy expression in the
Bloch basis $\{|\bk\nu\rangle\}$, reading~\cite{ama08}
\begin{equation}
\Sigma^{\hfill}_{\nu\nu'}(\bk,\omega)=\hspace*{-0.1cm}
\sum_{\bR,mm'}\hspace*{-0.1cm}\bar{P}^{\bR*}_{\nu m}(\bk) 
\left(\Sigma^{\rm \bR}_{mm'}(\omega)-\Sigma^{\rm dc}\right)
\bar{P}^{\bR}_{m'\nu'}(\bk)\;,\label{eq:sig}
\end{equation}
whereby here, $\nu\nu'$ are band indices, $mm'$ denote Cr($3d$) states, $\bR$ are
Cr sites and $\bar{P}$ are the projections between Bloch and local space. The
local self-energy correction to DFT on a given Cr site is given by the difference
between the DMFT impurity self-energy $\Sigma^{\rm \bR}_{mm'}(\omega)$ and the
double-counting term $\Sigma^{\rm dc}$. For the key effects, one may restrict the 
discussion to in-plane correlations, since most electron-electron scattering will 
happen in the relevant twodimensional (2D) subspace of the layered material. 
Figure~\ref{fig:sigma} displays real and imaginary part of the self-energy 
$\Sigma^{\rm 2D}_{\nu}(\omega)$, representing the diagonal elements of the $k$-summed 
$\Sigma^{\hfill}_{\nu\nu'}(\bk,\omega)$ with $k_z=0$. Besides the exptected 
large self-energy on the Cr-dominated bands, there is also a sizable self-energy
amplitude on the cPd band. On the contrary, for the bands further below in energy,
$\Sigma^{\rm 2D}_{\nu}(\omega)$ turns out very small.
 
In other words, albeit no explicit local Coulomb repulsion is taken into
account within DFT+DMFT on the Pd sites, the scattering with Mott-localized Cr
electrons transfers some correlations onto the most-relevant Pd-dominated
band. This will not only effect the transport properties~\cite{dao15,ars16} 
but also the exchange interaction between the Cr spins. Due to the half-filled 
scenario of the Cr$(3d)$ sites, kinetic exchange of antiferromagnetic kind is most 
natural. The revealed {\sl additional} coupling to the itinerant Pd electrons could 
cause an RKKY-like contribution. Understanding the intricate PdCrO$_2$ 
magnetism~\cite{tak09,tak10,bil15,ars16,gha17,le18}, 
which should be also partly rooted in the frustration on the triangular Cr 
sublattice, will ask for a detailed assessment of the different exchange 
contributions.

\subsection{Doping effects}
The question arises, if the hidden Mott-insulating state in PdCrO$_2$ could be
perturbed such that the correlation physics emerges more blatantly, perhaps
triggering other interesting physics. Various doped Mott insulators, such as
e.g. cuprates, serve as prominent examples for the latter. Thus doping may 
provoke new phenomena in the given delafossite, providing the unique opportunity of 
a doped Mott insulator {\sl within} a conducting environment. For instance, from basic 
considerations, electron- or hole-doped Cr sites might give rise to additional 
low-energy QP(-like) states with challenging strongly-correlated character. 
This could e.g. lead to an enhancement of the unconventional anomalous Hall signal
which was detected at stoichiometry~\cite{tak10}.
%%%%%%%%%%%%%%%%%%%%%%%%%%%%%%%%%%%%%%%%%%%%%%%%%%%%%%%%%%%%%%%%%%%%%%%%%%%%%%%%%%%%%%%%%
\begin{figure}[t]
(a)\hspace*{-0.2cm}\includegraphics*[width=8.25cm]{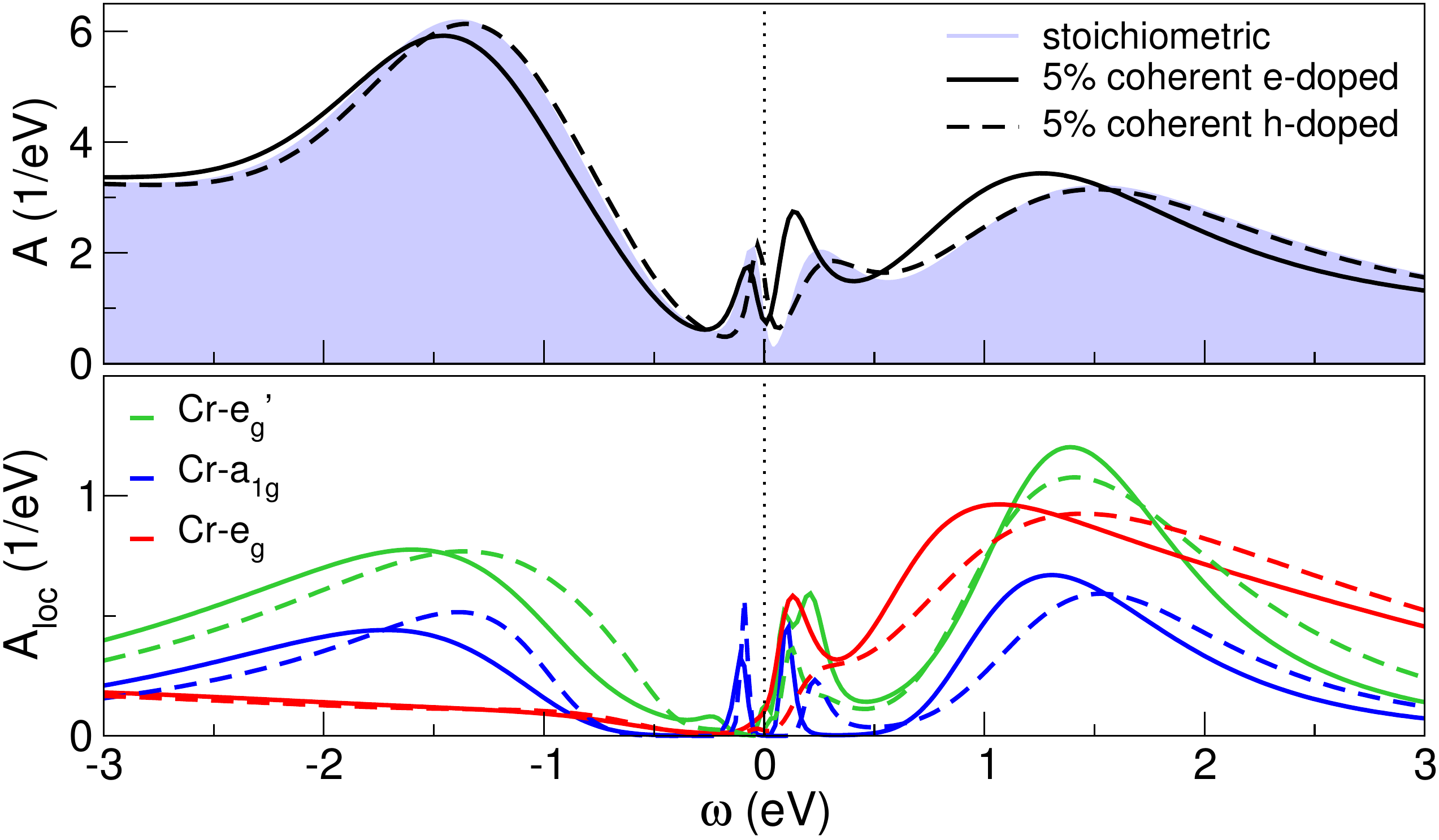}\\[0.2cm]
\includegraphics*[width=4.25cm]{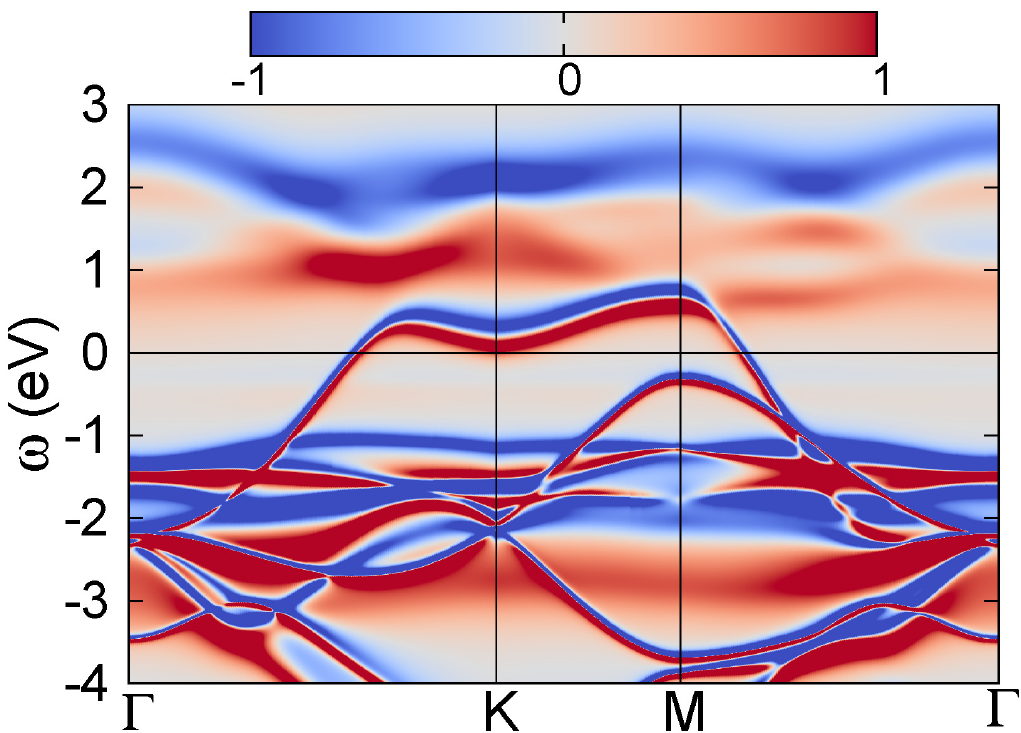}
\includegraphics*[width=4.25cm]{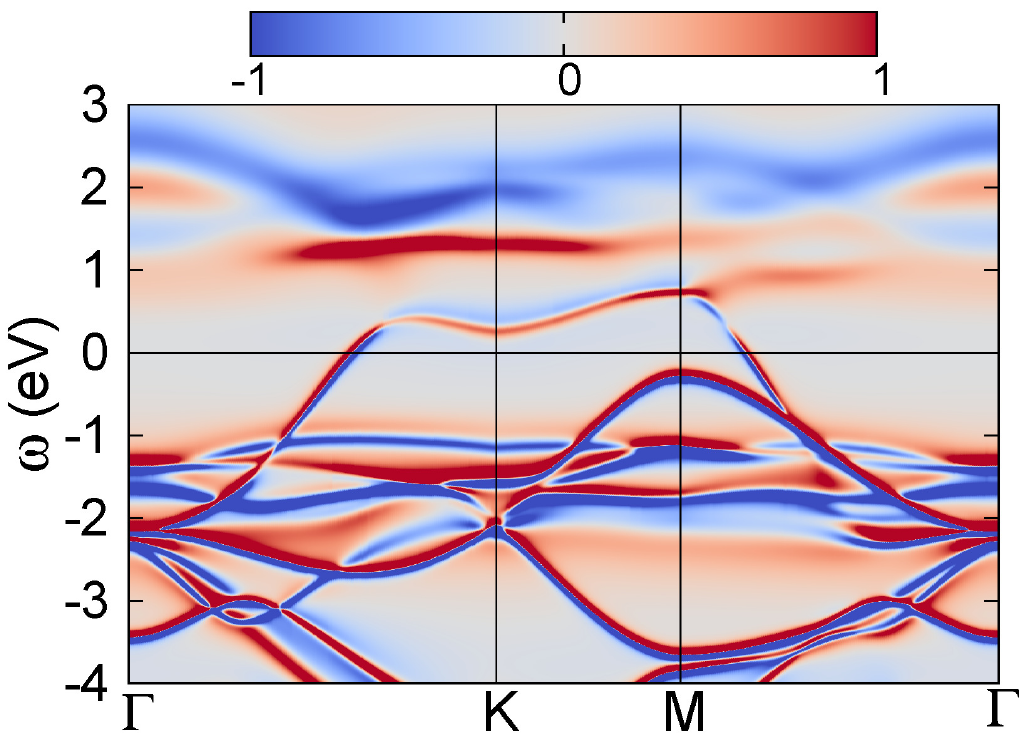}\\
\raggedright (b)
\caption{(color online) Electronic spectrum of PdCrO$_2$ with coherent doping of
5\% electrons or of 5\% holes, compared to the stoichiometric case.
(a) Integrated spectral functions, total (top) and local Cr (bottom).
(b) Difference between the angle-resolved doped spectrum and the stoichiometric
one, i.e. $A_{\rm e/h-doped}({\bf k},\omega)-A_{\rm undoped}({\bf k},\omega)$.
Left: electron doping, right: hole doping.}
\label{fig:vca}
\end{figure}
%%%%%%%%%%%%%%%%%%%%%%%%%%%%%%%%%%%%%%%%%%%%%%%%%%%%%%%%%%%%%%%%%%%%%%%%%%%%%%%%%%%%%%%%%

From the viewpoint of theory engineering, two ways of doping are feasible. 
First, through a perfectly-coherent charge doping by 
making use of the virtual-crystal approximation (VCA). A selected atomic species 
is thereby replaced by a pseudo atom of different nuclear charge $Z\pm\delta$. 
Second, and more realistic to experimental capabilities, through the introduction of 
explicit defects in a supercell approach.

\subsubsection{Coherent doping of electrons or holes}
Let's first deal with the somewhat more exemplary, but computationally simpler, coherent
doping within VCA. For a proof of principles, I focus on introducing electrons or holes 
via pseudizing Cr with a charge modification $\delta=0.05$, resembling 5\% electron(hole)
doping in terms of doped carriers per correlated site. Upon structural relaxation, the 
$z$ parameter of the crystal structure remains essentially unmodified with respect
to the stoichiometric case.

As expected, the small coherent dopings lead to a shift of the electronic spectrum, 
placing the maximum of the total QP peak further away(closer) to 
$\varepsilon^{\hfill}_{\rm F}$ for electron(hole) (see Fig.~\ref{fig:vca}a). In addition,
the spectral weight close to low energy based on the CrO$_2$ layers becomes somewhat
strenghtened with the doping. But a strong QP resonance from those layers still remains
absent. While the Cr-$a_{1g}$ contribution remains rather symmetric above and below
the Fermi level, the Cr-$e_{g}'$ and Cr-$e_{g}'$ orbitals show some carrier-doping 
characteristics for the case of electron doping.
The difference of the angle-resolved doped spectral functions and the one at 
stoichiometry shown in Fig.~\ref{fig:vca}b, shows that the cPd quasiparticle band 
shifts as expected with electron or hole doping. In the case of electron doping,
the conduction band at $K$ (carrying some Cr weight) touches the Fermi level. Thus a
Lifshitz-kind of Fermi-surface change occurs with already small electron doping.
%%%%%%%%%%%%%%%%%%%%%%%%%%%%%%%%%%%%%%%%%%%%%%%%%%%%%%%%%%%%%%%%%%%%%%%%%%%%%%%
\begin{table}[t]
\begin{ruledtabular}
\begin{tabular}{l|cccc}
doping                 & Cr-$e_g'$  & Cr-$a_{1g}$  & Cr-$e_g$ & total \\ \hline
stoichiometric         & 2.07    & 1.02      & 1.06  & 4.15  \\[0.1cm]
coherent 5\% electrons & 2.09    & 1.03      & 1.14  & 4.26  \\
coherent 5\% holes     & 2.07    & 1.02      & 1.08  & 4.17  \\
O vacancy              & 2.12    & 1.05      & 1.45  & 4.62 \\
Pd vacancy             & 2.06    & 1.02      & 1.15  & 4.23 \\
\end{tabular}
\end{ruledtabular}
\caption{Orbital-resolved Cr$(3d)$ occupations. For the explicit-defect cases,
averaging over all Cr sites in the supercell, respectively, is performed.}
\label{tab:explat}%
\end{table}
%%%%%%%%%%%%%%%%%%%%%%%%%%%%%%%%%%%%%%%%%%%%%%%%%%%%%%%%%%%%%%%%%%%%%%%%%%%%%%%

However, due to the small amount of doping and the coherent scenario, neglecting local 
structural distortions and symmetry breakings, the overall impact on the electronic 
spectrum remains without dramatic consequences. Interestingly, the original 
introduction of holes on the Cr sites on the atomic level, i.e. when initializing 
the calculation, results at DFT+DMFT convergence in a slightly {\sl larger} nominal 
Cr($3d$) charge (cf. Tab.~\ref{tab:explat}). Thus screening charge flow from nearby 
oxygens and/or the Pd layer overcompensates the initial Cr holes. 

\subsubsection{Doping by vacancies of oxygen or palladium type}
Doping closer to common experimental procedures may be realized by introducing
explicit defects in the system. I here choose vacancies of oxygen (O$_{\rm V}$)
and Pd (Pd$_{\rm V}$) type to proceed along these lines. For both doping cases,
2$\times$2$\times$1 supercells are constructed with structural optimization of 
the atomic positions. Both defect structures shelter 4 in-plane Pd, 4 in-plane Cr and
8 O sites, respectively. While the O$_{\rm V}$ disturbs the participating CrO$_6$ 
octahedra and drags the Pd ion above towards the CrO$_2$ plane, the Pd$_{\rm V}$
pulls the respective O ions above and below towards the Pd plane and thereby
distorts the corresponding CrO$_6$ octahedra. The present doping of 12.5\% 
O$_{\rm V}$s and of 25\% Pd$_{\rm V}$s is rather large, but suits the goal to 
study principle effects.

\begin{figure}[t]
\includegraphics*[width=8.5cm]{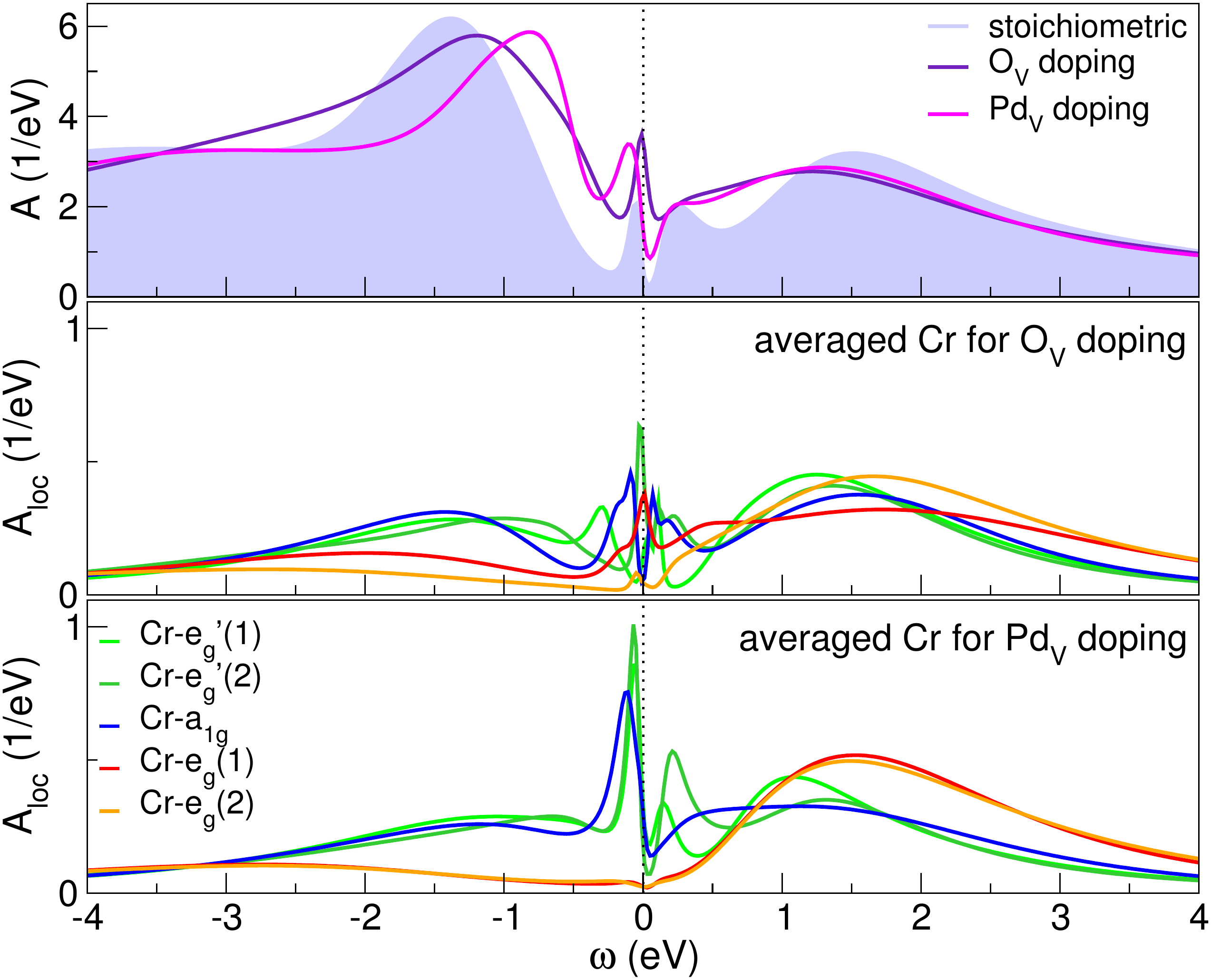}
\caption{(color online) Spectral properties of the O$_{\rm V}$-doped and 
Pd$_{\rm V}$-doped cases. Top: comparison of total spectral functions to 
stoichiometric case. Middle: Averaged Cr local spectral function for O$_{\rm V}$
doping. Bottom: Averaged Cr local spectral function for Pd$_{\rm V}$ doping.
}\label{fig:vacdop}
\end{figure}

The main spectral properties are summarized in Fig.~\ref{fig:vacdop}. Both doping
scenarios lead to an enhancement of the low-energy QP peak in the total spectral
function, which is obviously connected to the 'activation' of the originally
transport-inert CrO$_2$ planes. Indeed, the averaged local Cr spectra now show 
sizable weight close to $\varepsilon^{\hfill}_{\rm F}$. Importantly, whereas in
the previous case of coherent doping only the impact of charge alteration took
place, here, additional factors come into play. Besides the different total
electronic charge from taking out one O/Pd atom, the local symmetry breaking and 
structural distortions caused by this operation are also crucial for the doping
aspects. 
Concerning the charge doping on the Cr site, the orbital-resolved occupation of 
the Cr$(3d)$ states are given in Tab.~\ref{tab:explat}. As expected for early TM 
oxides~\cite{luo04,mit12,pav12,lec16}, introducing O$_{\rm V}$s dominatly 
adds charge to the TM-$e_g$ states, resulting here in a significantly 
larger nominal Cr charge. The associated Cr-$e_g(1)$ state (that is connected
to the vacant O site) develops furthermore sizable low-energy weight. On the other
hand, the Pd$_{\rm V}$ has only minor effects on the Cr$(3d)$ filling, with only
a slight increase of the Cr-$e_g$ electron count.
%%%%%%%%%%%%%%%%%%%%%%%%%%%%%%%%%%%%%%%%%%%%%%%%%%%%%%%%%%%%%%%%%%%%%%%%%%%%%%%%%%%%%%%%%
%%%%%%%%%%%%%%%%%%%%%%%%%%%%%%%%%%%%%%%%%%%%%%%%%%%%%%%%%%%%%%%%%%%%%%%%%%%%%%%%%%%%%%%%%

In comparison, the differences in charge doping of the Cr-$t_{2g}$ states between 
coherent and impurity doping are rather small. The still much larger response
in the latter case points to relevant doping contributions from distortions
and symmetry changes.
In the case of doping with oxygen vacancies, the Cr-based QP resonances at the 
Fermi level are sharp and the rather particle-hole symmetric local spectrum is 
reminiscent of a canonical strongly correlated system close to half-filled
Mott criticality~\cite{geo96} (cf. Fig.~\ref{fig:vacdop}, middle). Apparently, 
Pd$_{\rm V}$ doping tends to reduce the correlation strength in the CrO$_2$ planes 
by even stronger means (here also because of the larger nominal doping level). The 
local Cr spectral function with 25\% Pd$_{\rm V}$s has an appearance similar to 
a moderately correlated TM-$t_{2g}$ oxide~\cite{sek04}.

\section{Conclusions}
A detailed investigation of the correlated electronic structure of 
highly-conductive PdCrO$_2$ confirmed the hidden Mott insulator 
within the CrO$_2$ layers by theoretical means. 
The intriguing coexistence of a moderately correlated low-energy 
band and strongly correlated localized states renders the given material highly 
interesting, standing out of the large class of correlated TM oxides. Charge 
self-consistent DFT+DMFT is a proper tool to characterize the challenging 
electronic system, providing very good agreement with available spectral data
from experiment. The low-energy QP band observed in ARPES measurements is of 
dominant Pd$(4d)$ character, carrying subtle correlation effects from the 
coupling to the localized Cr$(3d)$ states. 
Impurity doping proves effective in 'melting' the hidden Mott insulator,
giving rise to intricate conductive correlated states. Here, it was shown that
in particular oxygen vacancies may be suitable to generate demanding correlated
transport. Notably, besides the sole carrier-doping impact, the effects of 
symmetry breaking and structural distortion appear also very important for the
doping impact.

So far nearly exclusively, experimental studies of PdCrO$_2$ focussed on
the highy-purity aspect and the magnetic properties at stoichiometry. Though
these facets are interesting in their own right, additional work on doped 
PdCrO$_2$ could be very exciting~\cite{maz17}. There is the chance 
for identifying electronic instabilities beyond the known low-temperature
antiferromagnetic(-like) order at stoichiometry. The present work dealt 
with the paramagnetic electronic structure at room temperature. A theoretical
study of the magnetic degrees of freedom, including e.g. a computation of exchange 
interactions, will be adressed in subsequent work. Moreover, revealing the
$T$-dependent multi-orbital lattice susceptibilities from a realistic many-body
perspective within DFT+DMFT~\cite{boe14} could shed further light on the intriguing 
electronic couplings in the system.

Finally, the PdCrO$_2$ compound (and also other delafossites) might be a promising
candidate as a building block for novel oxide heterostructures with underlying
triangular lattices. Note in this respect that the akin PdCoO$_2$ compound also
displays inert TM-oxide layers, yet not because of strong correlations
but due to band-insulating(-like) CoO$_2$ layers. Thus combining both compounds may
enable a controlled switching between Mott- and band-insulating characteristics.
In general, the unique layered structure, the wide range of electronically
different delafossite materials and the sensitivity to doping could open the possibility 
for engineering exceptional transport properties.

\begin{acknowledgments}
Financial support from the DFG LE-2446/4-1 project ``Design of strongly correlated
materials'' is acknowledged. Computations were performed at the 
JURECA Cluster of the J\"{u}lich Supercomputing Centre (JSC) under project 
number hhh08. 
\end{acknowledgments}

\bibliography{bibextra}

\end{document}